\documentclass[review,10pt]{fcs}
\usepackage{tabularray}
\usepackage{bm}
\title{Graph Neural Networks for Financial Fraud Detection: A Review}
\shorttitle{Graph Neural Networks for Financial Fraud Detection}
\author[1,2,3]{Dawei Cheng}
\author[1,3]{Yao Zou}
\author[4]{Sheng Xiang}
\author*[1,2,3]{Changjun Jiang}
\address[1]{Department of Computer Science and Technology, Tongji University, Shanghai 200092, China.}
\address[2]{Shanghai Artificial Intelligence Laboratory, Shanghai 200031, China.}
\address[3]{National Collaborative Innovation Center for Internet Financial Security, Shanghai 100045, China.}
\address[4]{AAII, University of Technology Sydney, Sydney, Australia 2007.}
\fcssetup{
  received       = {month dd, yyyy},
  accepted       = {month dd, yyyy},
  corr-email     = {cjjiang@tongji.edu.cn},
}

\begin{abstract}
The landscape of financial transactions has grown increasingly complex due to the expansion of global economic integration and advancements in information technology. This complexity poses greater challenges in detecting and managing financial fraud. This review explores the role of Graph Neural Networks (GNNs) in addressing these challenges by proposing a unified framework that categorizes existing GNN methodologies applied to financial fraud detection. Specifically, by examining a series of detailed research questions, this review delves into the suitability of GNNs for financial fraud detection, their deployment in real-world scenarios, and the design considerations that enhance their effectiveness. This review reveals that GNNs are exceptionally adept at capturing complex relational patterns and dynamics within financial networks, significantly outperforming traditional fraud detection methods. Unlike previous surveys that often overlook the specific potentials of GNNs or address them only superficially, our review provides a comprehensive, structured analysis, distinctly focusing on the multifaceted applications and deployments of GNNs in financial fraud detection. This review not only highlights the potential of GNNs to improve fraud detection mechanisms but also identifies current gaps and outlines future research directions to enhance their deployment in financial systems. Through a structured review of over 100 studies, this review paper contributes to the understanding of GNN applications in financial fraud detection, offering insights into their adaptability and potential integration strategies.

\end{abstract}


\begin{document}
\section{Introduction}
The rapid development of global economic integration and information technology in recent years has led to a significant increase in the scale and complexity of financial sector transactions. However, this expansion also brings about a broader range of financial fraud risks, resulting in a rise in criminal activities \cite{AlFalahi2019ConceptualBO,Mt2019THEEO}. According to a report by the Canadian Anti-Fraud Centre (CAFC), fraud losses in 2023 alone reached 554 million, marking a 4.3\% increase compared to the previous year, indicating an ongoing upward trend \cite{report1}. This data indicates that financial fraud not only causes direct economic losses to victims and businesses but also erodes public trust in the entire financial system \cite{MOTIE2024122156}. Therefore, it is crucial to build a financial fraud detection system to prevent unauthorized financial gain through illegal or deceptive means.

Traditional rule-based systems and classic machine learning methods have been used for fraud detection, but they often struggle with complex fraud patterns on large amounts of financial data \cite{ferreira2007detecting,Seeja2014FraudMinerAN,maes2002credit,ogwueleka2011data,gaikwad2014credit,ng2001discriminative,sahin2011detecting}. In contrast, deep learning has shown exceptional performance in various domains, leveraging its ability to learn intricate features from raw data \cite{DBLP:journals/corr/abs-2111-14818,feng2023layoutgpt,bai2021explainable,lopez2017deep}. Recent survey papers have demonstrated that deep learning models outperform traditional methods in terms of accuracy, adaptability, and scalability~\cite{Popat2018ASO,Ma2021ACS,ZhouXun2018}.
Some fraud detection models, such as those leveraging Graph Neural Networks (GNNs), have achieved remarkable progress and exhibited superior performance in the financial domain \cite{AHMED2016278,cheng2020graph,hamilton2017inductive,li2021temporal,DBLP:journals/corr/NiepertAK16}.
Utilizing deep learning techniques on graph, GNNs can ascertain advanced feature representations of nodes and edges, thereby identifying complex fraud patterns that may remain obscured within traditional data representations \cite{ma2021comprehensive}.
For example, analyzing transaction networks enables the detection of complex fraudulent activities such as money laundering, credit card fraud, and insurance fraud \cite{Alarab2020CompetenceOG,xia2022novel,sheu2022potential,xiang2023semi,liu2021graph,syeda2002parallel,ma2023fighting,ijcai2024p0830}.

Despite the potential of GNNs in fraud detection, there are practical challenges that need to be addressed. These include managing large-scale financial graph data and ensuring model adaptability to dynamic patterns \cite{Lin2020PaGraphSG,Tan2023QuiverSG,Park2022GinexSB,NEURIPS2022_2857242c,kazemi2020representation}. Additionally, transparency and interpretability of models are critical for enhancing the confidence of financial institutions \cite{ying2019gnnexplainer,wu2022discovering,sui2022causal}. To delve deeper into the existing challenges for GNNs in financial fraud detection, this study proposes the following detailed research questions (RQs):

\begin{itemize}
    \item \textbf{RQ1: How can various GNN methodologies in financial fraud detection be understood through a unified framework?}
    \item \textbf{RQ2: Why use GNNs for financial fraud detection? What roles do they play?}
    \item \textbf{RQ3: How to design GNNs suitable for financial fraud detection?}
    \item \textbf{RQ4: How are GNNs deployed in real-world financial fraud detection scenarios, and what are their impacts?}
    \item \textbf{RQ5: What are the current challenges and future directions for GNNs in financial fraud detection?}
\end{itemize}

To address these research questions, our contributions, summarized as follows, are designed to offer a holistic view of GNNs in financial fraud detection:

\begin{itemize}
\item \textbf{Establishing a Unified Framework:} We propose a comprehensive framework for categorizing and analyzing GNN methodologies in financial fraud detection, enhancing the clarity and depth of our understanding of this field. For a practical implementation of these methodologies, the AntiFraud (\url{{https://github.com/AI4Risk/antifraud}}) project provides an opensource example on GitHub, which demonstrates the application of these techniques.
\item \textbf{Exploring the Suitability of GNNs:} We provide a detailed examination of why GNNs are uniquely positioned to address financial fraud detection, emphasizing their intrinsic capabilities.
\item \textbf{Design Considerations for GNNs:} We offer insights into the nuanced design process of GNNs tailored for financial fraud detection, focusing on how architectural and strategic choices impact their effectiveness.
\item \textbf{Highlighting Real-world Applications:} We showcase examples of GNNs in action within the financial sector, discussing the outcomes and implications. Additional resources can be found on GitHub. (\small{\url{https://github.com/AI4Risk/awesome-graph-based-fraud-detection}})
\item \textbf{Charting Future Directions:} We explore the landscape of current challenges and future opportunities in GNN research for financial fraud detection, aiming to inspire further innovation in this field.
\end{itemize}

\noindent\textbf{Differences between this survey and existing ones.}  While there are numerous reviews that explore various machine learning techniques in financial fraud detection \cite{ashtiani2021intelligent,west2016intelligent,ngai2011application,al2021financial,yue2007review,sharma2012review,AHMED2016278,ali2022financial}, these generally focus on conventional methods and often overlook the potential of GNN. For instance, studies \cite{ashtiani2021intelligent,ali2022financial,al2021financial,yue2007review,AHMED2016278}  discuss a wide range of deep learning technologies applied in fraud detection but neglect GNNs, while \cite{ngai2011application} and \cite{kim2021systematic} address GNN applications in specific subdomains, without covering the financial fraud detection field comprehensively.
Moreover, although some surveys \cite{MOTIE2024122156} briefly introduce systems based on graph learning, they merely touch upon the application of GNNs in financial fraud detection and do not delve deep. Another review \cite{ZHOU20205}, attempts to categorize GNN applications from multiple perspectives but focuses more on the technical framework rather than their practical utility in detecting financial fraud. These publications do not discuss in depth the advantages and limitations of GNN technologies.

\begin{figure*}
    \centering
    \includegraphics[width=1\textwidth]{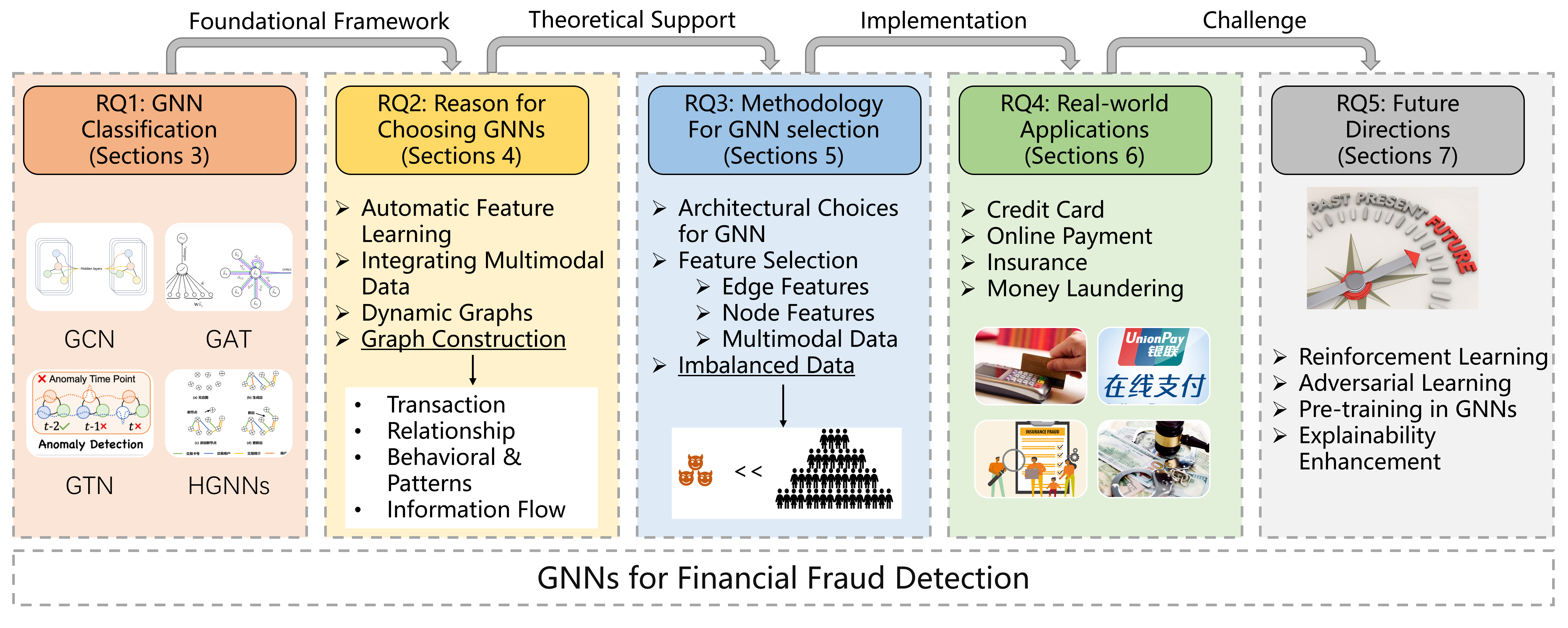}
    \caption{An overview of the road map of this paper. We first introduce the mainly utilized graph neural networks (GNNs) for financial fraud detection. Secondly, we summarize the reasons for choosing GNNs. After that, we introduce the methods and special tricks in GNN architecture design for financial fraud detection tasks. Then we go through the real-world applications and future directions for GNN-based financial fraud detection.}
    \label{fig:Taxonomy}
\end{figure*}

This review not only summarizes the existing literature but also explores the advantages of GNN technologies, relevant architectural choices, current challenges, and future directions. By classifying and examining over 100 relevant studies, this paper aims to provide a detailed reference for further research and application of GNNs in the domain of financial fraud detection. This review paper points out that future research is anticipated to concentrate on augmenting the scalability, adaptability, and interpretability of GNNs, exploring cross-domain methodologies and multimodal data integration to bolster financial fraud detection and ensure financial security. Figure \ref{fig:Taxonomy} represents the road map of the rest of the paper. We introduced the main GNN methods in section~\ref{sec:gnn_classi}. Then sections~\ref{sec:fin_data} and~\ref{sec:feat_eng} introduce the reason to choose GNNs and GNN design in the financial field for GNNs, respectively.

\begin{table}
\centering
\caption{The summary of notations in this paper.}
\tabcolsep 28pt
\begin{tabular}{cc}
\toprule
\textbf{Symbol} & \textbf{Description} \\
\midrule
$G$ & Graph \\
$V$ & Set of nodes \\
$E$ & Set of edges \\
$n$ & Number of nodes \\
$\bm{I}_n \in \mathbb{R}^{n \times n}$ & Identity matrix \\
$\bm{A} \in \mathbb{R}^{n \times n}$ & Adjacency matrix \\
$\bm{D} \in \mathbb{R}^{n \times n}$ & Degree matrix (diagonal) \\
$\bm{L} \in \mathbb{R}^{n \times n}$ & Laplacian matrix \\
$\bm{U} \in \mathbb{R}^{n \times n}$ & Eigenvector matrix \\
$\bm{\Lambda} \in \mathbb{R}^{n \times n}$ & Eigenvalue matrix (diagonal) \\
$\bm{u}_i \in \mathbb{R}^n$ & $i$-th eigenvector \\
$\bm{X} \in \mathbb{R}^{n \times D}$ & Node feature matrix \\
$\bm{x}_i \in \mathbb{R}^D$ & Features of the $i$-th node \\
$\bm{f} \in \mathbb{R}^n$ & Signal \\
\bottomrule
\end{tabular}
\label{table:symbols}
\end{table}
\section{Preliminaries}
\subsection{Problem Definitions}
In this section, we first provide the definitions of the common notations used, as shown in Table~\ref{table:symbols}. We define graph \(G = (V, E, \bm{A})\) as an undirected graph, where \(V\) is the set of nodes, \(|V| = n\) is the number of nodes, \(E\) is the set of edges, and \(\bm{A} \in \mathbb{R}^{n \times n}\) is the adjacency matrix; the entry \(\bm{A}_{i,j}\) of the matrix is 1 if there is an edge between nodes \(i\) and \(j\), and 0 otherwise. The degree matrix \(\bm{D} \in \mathbb{R}^{n \times n}\) is a diagonal matrix where each entry \(\bm{D}_{i,i}\) is the sum of the \(i\)-th row of \(\bm{A}\), \(\sum_j \bm{A}_{i,j}\). The graph Laplacian matrix is defined as \(\bm{L} = \bm{I}_n - \bm{D}^{-\frac{1}{2}} \bm{A} \bm{D}^{-\frac{1}{2}}\), where \(\bm{I}_n \in \mathbb{R}^{n \times n}\) is the identity matrix. The symmetric normalized Laplacian matrix is another form of the Laplacian matrix, defined as \(\bm{L} = \bm{U} \bm{\Lambda} \bm{U}^\top\). Here \(\bm{U} = \{\bm{u}_i | i = 1, \ldots, n\}\) is the matrix of eigenvectors, and \(\bm{\Lambda} = \text{diag}(\{\lambda_i | i = 1, \ldots, n\})\) is the diagonal matrix of eigenvalues, and \(\bm{u}_i\) is the \(i\)-th eigenvector. We define \(\bm{X} \in \mathbb{R}^{n \times D}\) as the node feature matrix of graph \(G\), where \(\bm{x}_i \in \mathbb{R}^D\) is the feature vector of node \(i\). Then we introduce some extended definitions as follows:\\
\noindent\textbf{Homogeneous and Heterogeneous Graphs.} A graph \(G = (V, E)\) can be extended to a heterogeneous graph defined as \(G = (V, E, \phi, \psi)\), where \(\phi: V \to \mathcal{A}\) assigns types to nodes and \(\psi: E \to \mathcal{R}\) assigns types to edges. A graph is \textit{homogeneous} if \(\phi\) and \(\psi\) each map to a single type; otherwise, it is \textit{heterogeneous}.

\noindent\textbf{Multi-relation graph.} A Multi-relation graph is a graph where edges have different types.

\noindent\textbf{Dynamic graph.} A dynamic graph is defined as a sequence of graphs \(G^{seq} = \{G_1, \ldots, G_T\}\), where \(G_i = (V_i, E_i)\), for \(i = 1, \ldots, T\), where \(V_i,E_i\) are the set of nodes and edges for the \(i\)-th graph in the sequence respectively.

\subsection{Graph Neural Networks}
Typically, GNNs employ a message passing mechanism that incrementally aggregates information from the neighborhoods. Specifically, the message passing process at the \(k\)-th layer in a GNN is delineated in two primary phases:
\begin{equation}
\bm{m}^{(k)}_{i} = \text{AGGREGATE}\left(\{\bm{h}^{(k-1)}_{j} : j \in \mathcal{N}(i)\}\right)
\end{equation}
\begin{equation}
\bm{h}^{(k)}_{i} = \text{UPDATE}\left(\bm{h}^{(k-1)}_{i}, \bm{m}^{(k)}_{i}, \bm{x}_{i}\right)
\end{equation}
where \(\mathcal{N}(i)\) denotes the set of neighboring nodes of node \(i\), AGGREGATE represents a permutation invariant aggregation function, and \(\bm{x}_{i}\) is the feature vector of node \(i\) from the node feature matrix \(\bm{X}\). Following \(K\) iterations of message passing, the resultant node embeddings \(\bm{H}^{(K)}\) are then employed to execute the designated task. Figure \ref{fig:node-edge-graph} illustrates the tasks of GNNs at the node, edge, and graph levels, and Table \ref{tab:different-level-tasks} provides a comprehensive summary of the tasks associated with each of these levels. We briefly introduce the tasks of GNNs at three different levels as follows:

\begin{figure}
    \centering
   \includegraphics[width=0.5\textwidth]{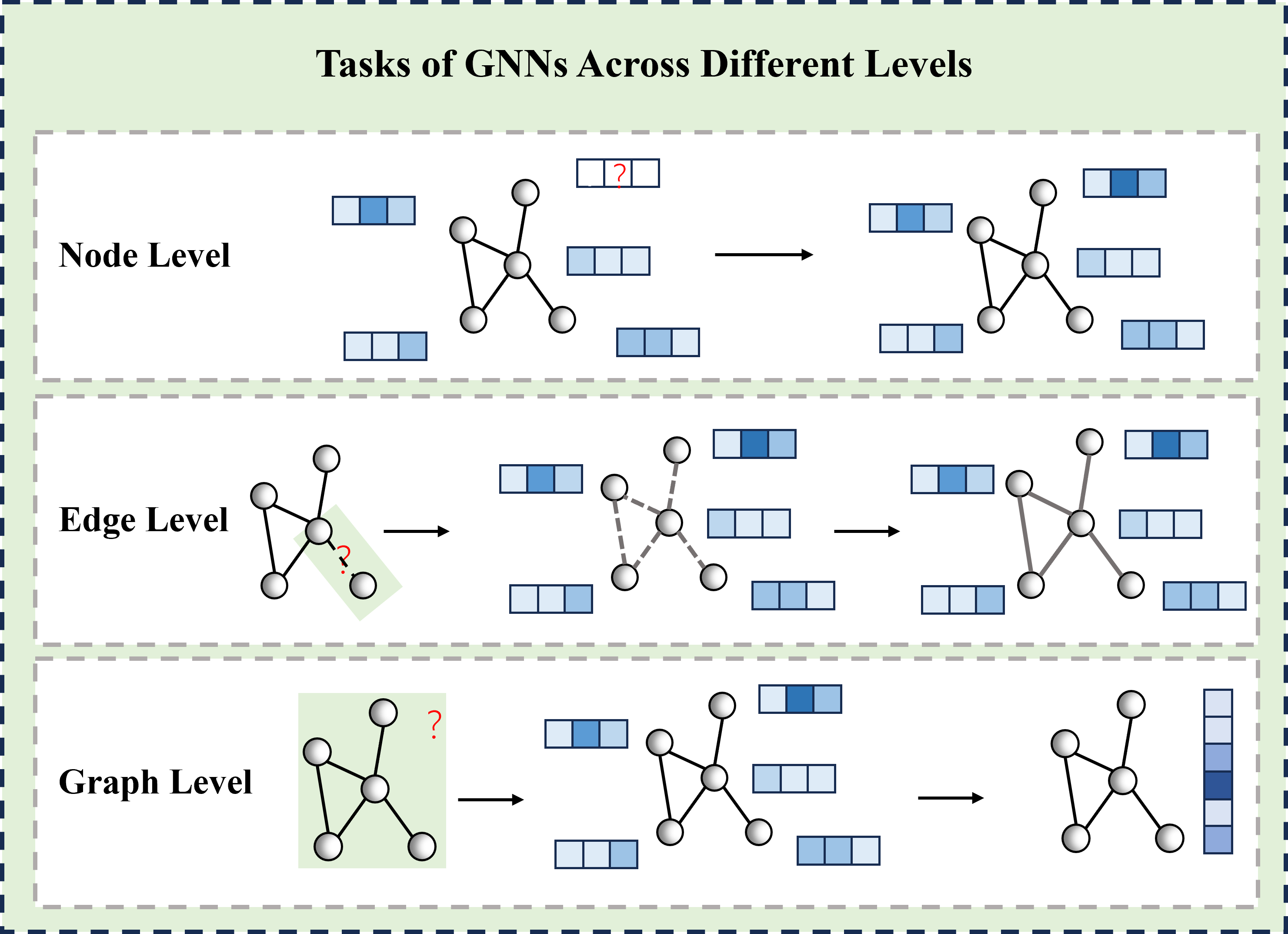}
    \caption{The tasks of GNNs at different levels. For node level, the label of each node is determined by its features and neighbors. For edge level, the label of each edge is determined by the features of source \& target node. For graph level, the property is determined by the features of all nodes or edges.}
    \label{fig:node-edge-graph}
\end{figure}

\begin{table*}
\centering
\caption{Summary of tasks across different levels in financial fraud detection.}
\label{tab:different-level-tasks}
\resizebox{\linewidth}{!}{%
\begin{tblr}{
  hline{1,5} = {-}{0.08em},
  hline{2} = {-}{0.05em},
}
\textbf{Level} & \textbf{Graph Tasks}                       & \textbf{Application Scenarios}                         \\
Node      & Node Classification                        & Credit Card Fraud Detection, Insurance Fraud Detection, etc. \\
Edge      & Edge Classification, Link Prediction       & Fraudulent Transaction Prediction, etc.              \\
Graph    & Graph Classification, Attribute Prediction & Systemic Risk Assessment, etc.                   \\
\end{tblr}
}
\end{table*}

\noindent\textbf{Node-Level:} GNNs are instrumental in node classification for financial networks, determining the label of each node based on its attributes and relationships. They effectively classify nodes for detecting credit card \cite{xiang2023semi,liu2021graph,syeda2002parallel,ijcai2024p839} and insurance fraud \cite{ma2023fighting,Zhang_Cheng_Yang_Ouyang_Wu_Zheng_Jiang_2024} by analyzing transaction patterns and claimant-provider relationships within financial networks.

\noindent\textbf{Edge-Level:} At the edge level, GNNs are utilized for tasks such as edge classification and link prediction, which are pivotal for predicting fraudulent transactions in finance \cite{zhang2022efraudcom,zheng2023midlg}.

\noindent\textbf{Graph-Level:} At the graph level, GNNs address challenges in graph classification and attribute prediction and are applicable in finance for systemic risk assessment \cite{balmaseda_predicting_2023}.

\section{Graph Neural Networks in Financial Fraud Detection}
\label{sec:gnn_classi}
This survey provides an in-depth exploration of GNNs applied in the domain of financial fraud detection, a critical area demanding advanced analytical techniques due to its complex and dynamic nature. Over the past five years, we have rigorously reviewed more than 100 high-quality papers from prestigious conferences and journals. This section categorizes the GNNs in financial fraud detection into four main types, reflecting the evolving landscape of graph-based deep learning technologies.

\subsection{Graph Convolutional Networks}
Graph Convolutional Networks (GCNs) process data structured in graph form, effectively identifying cross-transaction fraudulent activity patterns within transaction networks. The graph convolution operation on a graph \(G = (V, E, \bm{A})\) is formalized as:
\begin{align}
\bm{H}^{(k+1)} &= \sigma(\bm{\hat{D}}^{-\frac{1}{2}} \bm{\hat{A}} \bm{\hat{D}}^{-\frac{1}{2}} \bm{H}^{(k)} \bm{W}^{(k)})
\end{align}
where \(\bm{\hat{A}} = \bm{A} + \bm{I}_n\) and \(\bm{\hat{D}}\) is the degree matrix of \(\bm{\hat{A}}\). The modified adjacency matrix $ \bm{\hat{A}} $ ensures that each node has a self-loop, which is crucial for capturing the node's own information in its feature updates. The normalization step ensures that the influence of each neighbor is balanced, preventing any single node's impact from dominating the feature updates. This highlights GCNs' capability in uncovering subtle irregularities in financial transactions.

GCNs efficiently learn features and recognize patterns in financial data by capturing local connectivity and automatically discerning node and edge features. However, GCNs can suffer from over-smoothing, leading to the loss of distinct node characteristics, and are prone to overfitting with limited data.

\subsection{Graph Attention Networks}
Graph Attention Networks (GATs) employ an attention mechanism, allowing the model to focus on the most relevant parts of the graph. The attention coefficients can be computed as:
\begin{align}
\alpha_{ij} &= \frac{\exp(\text{LeakyReLU}(\bm{a}^\top [\bm{W} \bm{h}_i \| \bm{W} \bm{h}_j]))}{\sum_{k \in \mathcal{N}(i)} \exp(\text{LeakyReLU}(\bm{a}^\top [\bm{W} \bm{h}_i \| \bm{W} \bm{h}_k]))}
\end{align}
where \(\alpha_{ij}\) represents the importance of node \(j\)'s features to node \(i\), $\cdot^\top$ denotes transposition and $\|$ is the concatenation operation. The updated features for a node \(i\) are then computed as a weighted sum of the features of its neighbors:
\begin{align}
\bm{h}_i' = \sigma\left(\sum_{j \in \mathcal{N}(i)} \alpha_{ij} \bm{W} \bm{h}_j\right)
\end{align}
To stabilize the learning process of self-attention, multi-head attention mechanism is applied, resulting in the following output feature representation:
\begin{align}
 \bm{h}_{i}^{\prime}=\|_{k=1}^{K} \sigma\left(\sum_{j \in \mathcal{N}_{(i)}} \alpha_{i j}^{k} \bm{W}^{k} \bm{h}_{j}\right)
\end{align}
where $\alpha_{ij}^k$ are normalized attention coefficients computed by the $k$-th attention mechanism, and $\bm{W}^k$ is the corresponding input linear transformation’s weight matrix.

Specially, if we perform multi-head attention on the final layer of the network, it can be mathematically formulated as:
\begin{align}
 \bm{h}_{i}^{\prime}= \sigma\left( \frac{1}{K} \sum_{k=1}^{K} \sum_{j \in \mathcal{N}_{(i)}} \alpha_{i j}^{k} \bm{W}^{k} \bm{h}_{j}\right)
\end{align}

The attention mechanism is particularly useful in applications such as financial fraud detection \cite{veličković2018graphattentionnetworks}. GATs outperform GCNs by dynamically allocating attention weights to neighbors, enabling adaptive focus on relevant nodes for improved complex pattern detection in financial fraud.

\subsection{Graph Temporal Networks}
Graph Temporal Networks (GTNs) incorporate the dynamic changes in financial transactions, utilizing expressions like:
\begin{align}
\bm{H}^{(t+1)} &= \sigma(\bm{A}^{(t)} \bm{H}^{(t)} \bm{W}^{(t)})
\end{align}
to capture dynamic transactional data. These networks are particularly effective in identifying fraud within datasets sensitive to temporal patterns, such as credit card transactions and high-frequency trading \cite{shixu_wang2021temporal,shixu_huang2023spatio}. To enhance the model's sensitivity to temporal dynamics, a temporal attention mechanism can be introduced, enabling the model to focus on the most significant parts of the graph at different times. For example, a temporal attention mechanism can be formulated as:
\begin{gather}
\alpha^{(t)} = \text{softmax}\left(\bm{q}^\top \tanh(\bm{W}_t \bm{H}^{(t)} + \bm{b}_t)\right), \\
\bm{H}_{\text{att}}^{(t)} = \sum_{t}\alpha^{(t)} \bm{H}^{(t)}
\end{gather}
where \(\alpha^{(t)}\) are the attention weights at time \(t\), determined by the learnable parameters \(\bm{q}\), \(\bm{W}_t\), and \(\bm{b}_t\). The vector \(\bm{q}\) is a query vector that projects the transformed node features into a scalar, thereby computing a preliminary attention score for each node. This score indicates the relevance or importance of each node's features at time \(t\) before normalization by the softmax function.
Additionally, to accommodate multiple time steps and interactions between them, recurrent mechanisms can be employed:
\begin{align}
\bm{H}^{(t)} &= \text{GRU}(\bm{H}^{(t-1)}, \bm{X}^{(t)})
\end{align}
where \(\bm{H}^{(t)}\) and \(\bm{H}^{(t-1)}\) are the hidden states of the nodes at time \(t\) and \(t-1\), respectively, \(\bm{X}^{(t)}\) represents the input features at time \(t\), and \(\text{GRU}\) denotes a gated recurrent unit used to integrate temporal information \cite{jiang2021mafi}. This addition of a recurrent layer allows the model to maintain a memory of past states, thus improving its ability to discern time-sensitive patterns that static GCNs cannot.

\subsection{Heterogeneous Graph Neural Networks}
Heterogeneous Graph Neural Networks (HGNNs) are designed to handle graphs composed of different types of nodes and edges, enabling the model to capture a richer set of relationships and properties within the data. To process heterogeneous graphs, it often employs a type-specific transformation mechanism, which can be mathematically formulated as:
\begin{align}
\bm{H}^{(k+1)}_a &= \sigma\left(\sum_{r \in \mathcal{R}} \sum_{b \in \mathcal{A}} \bm{W}^{(k)}_{r, a, b} \cdot \text{AGG}\left(\bm{H}^{(k)}_b, \bm{E}_{r, a, b}\right)\right)
\end{align}
where \(\bm{H}^{(k)}_a\) represents the node features of type \(a\) at layer \(k\), \(\bm{W}^{(k)}_{r, a, b}\) is the weight matrix for relation \(r\) between node types \(a\) and \(b\), \(\bm{E}_{r, a, b}\) denotes the edges of type \(r\) between node types \(a\) and \(b\), and \(\text{AGG}(\cdot)\) is an aggregation function that combines features from neighboring nodes. The function \(\sigma\) is a non-linear activation. Common choices for \(\text{AGG}\) include sum, mean, and attention mechanisms, each providing different ways to emphasize the contributions of neighbors.

To enhance the capabilities of heterogeneous networks, the meta-path based approach \cite{liang2022meta,wang2019heterogeneous,meng2015discovering} can be employed:
\begin{align}
\bm{H}^{(k+1)}_{\text{meta}} &= \sigma\left(\sum_{p \in \mathcal{P}} \bm{W}^{(k)}_p \cdot \text{AGG}_{p}\left(\{\bm{H}^{(k)}_b \mid b \in \text{path}(p)\}\right)\right)
\end{align}
where \(\bm{H}^{(k+1)}_{\text{meta}}\) indicates the updated node features along a specific meta-path \(p\), \(\bm{W}^{(k)}_p\) represents the weight matrix specific to meta-path \(p\), and \(\text{AGG}_{p}\) is an aggregation function tailored to the meta-path \(p\) that considers node types and relations along the path. Meta-paths enable more complex relationships and structures to be utilized effectively, integrating context from multiple types of interactions.

Moreover, incorporating an attention mechanism \cite{liu2021intention,ghosh2023gosage,wang2021modeling} can dynamically weight the importance of different types of relationships:
\begin{align}
\alpha_{r, a, b} &= \text{softmax}\left(\bm{q}_{r, a, b}^\top \sigma(\bm{W}_{r, a, b} \cdot \bm{H}^{(k)}_b)\right), \\
\bm{H}^{(k+1)}_a &= \sum_{r \in \mathcal{R}} \sum_{b \in \mathcal{A}} \alpha_{r, a, b} \cdot \bm{W}^{(k)}_{r, a, b} \cdot \bm{H}^{(k)}_b
\end{align}
where \(\alpha_{r, a, b}\) are attention weights that assess the relative importance of each neighboring node \(b\) of type \(b\) via relation \(r\) to node \(a\), allowing the network to focus adaptively on the most informative parts of the heterogeneous structure.

\subsection{GNN Application Overview}
In exploring the down stream tasks of various GNNs for detecting financial fraud, Table \ref{tab:gnn_applications} offers a comprehensive summarization. This table compiles a range of GNN approaches, including GCN, GAT, GTN, and HGNN, to illustrate their usage in financial fraud detection. Each method is collected based on its implementation details and characteristics, aiming to provide us with a clear perspective on which GNN strategies are effective in combating financial fraud.

\textbf{GCNs} such as Skip-GCN \cite{gcn_DBLP:journals/corr/abs-1908-02591} and EWS-GCN \cite{gcn_9338368} are adept at leveraging node and edge features but may fall short in capturing the temporal dynamics of fraud patterns. \textbf{GATs} like ASA-GNN \cite{tian2023asa} and ABGRL \cite{sun2024adaptive}, offer adaptive attention mechanisms that enhance the detection of nuanced fraudulent behaviors, yet they can be computationally intensive. \textbf{GTNs} such as STAGN \cite{cheng2020graph} and STGN \cite{xie2023spatial}, proficient in detecting time-sensitive fraud patterns, require more computational overhead due to their temporal attention modules. \textbf{HGNNs}, including LIFE \cite{gcn_li2021live} and MultiFraud \cite{wu2024heterogeneous}, excel at integrating heterogeneous information, providing a rich context for fraud detection, but their complexity can be a double-edged sword, demanding more resources and making them less suitable for scalability-focused applications.

Despite their efficacy, GNNs encounter significant scalability challenges when scaling up to large financial networks. The high memory and computational demands on a single machine primarily stem from GPU memory constraints and the extensive time required for matrix multiplication operations. Additionally, in distributed environments, the high communication costs related to obtaining and combining graph embeddings of nodes' multi-hop neighbors in each training batch hinder the speed benefits of parallelization \cite{DBLP:journals/corr/abs-2111-14818}. These challenges highlight the need for more efficient GNN architectures to handle larger financial networks with quick response times for real-time fraud detection.

\begin{table*}
\centering
\caption{Summary of GNN methods for financial fraud detection.}
\label{tab:gnn_applications}
\resizebox{\linewidth}{!}{%
\begin{tblr}{
  hline{1,15} = {-}{0.08em},
  hline{2} = {-}{0.05em},
}
\textbf{Approach} & \textbf{Category} & \textbf{Key Idea}                           
& \textbf{Level} & \textbf{Application}           \\
Skip-GCN (2019) \cite{gcn_DBLP:journals/corr/abs-1908-02591}  & GCN               & Skip Connection                             
& Node Level     & Anti-Money Laundering          \\
EWS-GCN (2020) \cite{gcn_9338368}   & GCN               & Edge Weight-shared Convolution              
& Node Level     & Credit Scoring                 \\
FMGAD (2023) \cite{gcn_xu2023few}     & GCN               & Deep-GNN                                    
& Node Level     & Credit Card Fraud Detection    \\
Li et al. (2021) \cite{li2021fraud} & GAT               & Self-Attention Mechanism                    
& Node Level     & Credit Card Fraud Detection    \\
ASA-GNN (2023) \cite{tian2023asa}   & GAT               & Adaptive Sampling and Aggregation           
& Node Level     & Transaction Fraud Detection    \\
ABGRL (2024) \cite{sun2024adaptive}     & GAT               & Adaptive Attention Convolution              
& Node Level     & Phishing Accounts Detection    \\
STAGN (2020) \cite{cheng2020graph}     & GTN               & Spatial-Temporal Attention Module           
& Node Level     & Credit Card Fraud Detection    \\
STGN (2023) \cite{xie2023spatial}      & GTN               & Spatial-Temporal Attention Module           
& Node Level     & Credit Card Fraud Detection    \\
LIFE (2021) \cite{gcn_li2021live}      & HGNN              & Heterogeneous Graph Representation Learning 
& Node Level     & Online Payment Fraud Detection \\
C-FATH (2021) \cite{wang2021modeling}    & HGNN              & Community-based Filtering                   
& Node Level     & Online Payment Fraud Detection \\
CGNN (2022) \cite{zhang2022efraudcom}      & HGNN              & Competitive Mechanism                      
& Edge Level     & E-commerce Fraud Detection     \\
GoSage (2023) \cite{ghosh2023gosage}    & HGNN              & Node Level and Relation Level Attention     
& Node Level     & Collusion Fraud Detection      \\
MultiFraud (2024) \cite{wu2024heterogeneous}& HGNN              & Multi-view Representation Learning         
& Node Level     & Supply Chain Fraud Detection
\end{tblr}
}
\end{table*}

\section{Reasons to Choose GNNs for Financial Fraud Detection}
\label{sec:fin_data}
GNNs have shown promising results in the field of financial fraud detection by leveraging their unique abilities to model complex relationships, learn features automatically, handle dynamic graphs, and integrate multimodal data. This section explores these aspects in detail, highlighting the versatility and effectiveness of GNNs in tackling the challenges associated with detecting fraudulent activities in financial systems.

\subsection{Graph Construction from Financial Data}
To address the requirements of fraud detection within the financial sector, constructing financial-oriented graphs is essential for uncovering \emph{abnormal transaction patterns}, \emph{unusual behavioral interactions}, and \emph{potential fraud clusters}. This subsection introduces various methods for constructing graphs based on financial data, aiming to facilitate the detection and prediction of fraudulent behavior:

\noindent\textbf{Transaction Graph Construction.} Transaction data is pivotal for revealing fraudulent activities. Creating a transaction network (graph), with nodes representing individual accounts or entities (such as individuals, companies) and edges representing transactions (e.g., transfers, payments) \cite{altman2024realistic,singh2019anti}, enables the effective identification of abnormal transaction patterns \cite{AdaptiveGNN2023}. For example, frequent small transactions or large funds flows within a short period may indicate fraudulent activity \cite{reurink2019financial}. Analyzing the network's topology, such as node centrality and community divisions, can help identify fraud rings or key fraudulent accounts.

\noindent\textbf{Relationship-Based Graph Construction.} Beyond direct transaction data, the relationships between financial entities are crucial for fraud detection \cite{shi2023enhancing}. Constructing complex network (graph) that include various entities (like individuals, companies, bank accounts) and their relationships (such as ownership, control relations) can reveal the cooperation networks behind complex fraudulent activities like money laundering or insider trading\cite{d2016complex}. For example, Mao and others utilize the inter-enterprise relationships to establish a network \cite{mao2022using}.

\noindent\textbf{Behavioral and Device Pattern Graph.} Constructing graphs based on user behaviors and device usage patterns (e.g., logins, queries, transaction requests, device types, operating systems, and device IDs) can reveal the similarities or abnormal patterns between different accounts \cite{akoglu2015graph,Egele2013COMPA,zou2024subgraph}. By connecting accounts whose behavior and device usage similarity exceeds a certain threshold, it's possible to uncover networks of accounts potentially operated by fraudsters. This includes identifying users who switch devices in patterns that deviate from typical user behavior, or who use devices known to be associated with fraudulent activities \cite{jiang2021mafi}. Such graphs significantly enhance machine learning models' ability to recognize atypical behavior and device usage patterns, thereby predicting and identifying potential fraud more effectively \cite{ma2021comprehensive}.

\noindent\textbf{Information Flow Graph Construction.}
Malicious actors often exploit misleading information for market manipulation. Constructing information flow graphs \cite{giudici2016graphical} by analyzing the relationship between public information (e.g., news, announcements, social media discussions) and financial market activities can help identify patterns of information manipulation and the centers of influence, thereby detecting related fraudulent activities \cite{bright2022,qgnn2023,kurshan2021graph}.

These methods illustrate the diverse approaches to graph construction from financial data, each contributing uniquely to the identification and prediction of fraud.

\subsection{Automatic Feature Learning}
GNNs greatly enhance financial fraud detection through their ability to learn features automatically, unlike traditional machine learning methods that require extensive feature engineering. GNNs extract relevant features from graph data, which is crucial for keeping pace with the dynamic nature of financial fraud. The foundational layer of GNNs, the message passing mechanism, enables a sophisticated form of feature extraction. It integrates attributes from both nodes and edges, which are rich in data that can signal fraudulent activity. This mechanism effectively combines the influences of neighboring nodes and updates each node's features through successive layers. This iterative process allows GNNs to reveal complex and non-linear relationships within the data, typical of fraudulent activities \cite{hamilton2017inductive}. The capacity for automatic feature learning---requiring no specific programming for feature identification---makes GNNs exceptionally effective at uncovering both sophisticated and previously undetected fraud schemes. Overall, the intrinsic feature learning of GNNs not only eases model development and training but also ensures rapid adaptation to new fraudulent tactics, playing a vital role in fighting financial crime.
\subsection{Dynamic Graphs}
Financial networks are inherently dynamic, continuously evolving with the addition of new transactions, accounts, and users. This dynamic nature presents significant challenges for fraud detection systems, which must adapt to effectively detect emerging fraudulent schemes.

\noindent\textbf{Challenges Posed by Dynamic Financial Graphs.}
The ever-changing structure of financial graphs complicates the task of fraud detection. As new nodes and edges are added, previously learned patterns may no longer apply, and new patterns of fraud can emerge. Traditional static models struggle to capture these evolving patterns, often leading to outdated or inaccurate fraud detection.

GNNs offer a promising solution to the challenge of modeling dynamic financial graphs. By incorporating the temporal aspect into their architecture, GNNs can update their node and edge representations in response to new information. Several approaches enhance GNNs' adaptability to dynamic graphs as follow:

\noindent\emph{Temporal Attention Mechanisms} allow GNNs to weigh the importance of different transactions or behaviors over time, focusing on those most indicative of fraud. This is particularly effective in emphasizing recent transactions that may signal current fraudulent activities.

\noindent\emph{Time-Decay Functions} \cite{kumar2019predicting} in the aggregation process enable GNNs to prioritize recent interactions, formalized as:
\begin{equation}
\bm{m}^{(k)}_i(t) = \text{AGGREGATE}\left(\{\alpha(t, t_j) \bm{h}^{(k-1)}_{j}(t_j) : j \in \mathcal{N}(i)\}\right)
\end{equation}
\begin{equation}
\bm{h}^{(k)}_{i}(t) = \text{UPDATE}\left(\bm{h}^{(k-1)}_{i}(t), \bm{m}^{(k)}_i(t)\right)
\end{equation}
where \(\alpha(t, t_j)\) diminishes the influence of older interactions over time, emphasizing more recent data. Unlike GRUs, which dynamically process temporal sequences, time-decay functions statically modulate past data's impact, ideal for prioritizing recent interactions.

\noindent\emph{Graph Temporal Networks} \cite{zhao2019t} combine the spatial structure captured by GCNs with the temporal dynamics processed by RNNs, offering a robust framework for dynamic graph analysis. The specific details can be found in Section 3.3.

\noindent\emph{Dynamic Embeddings.} \cite{zhou2020dynamic} update node representations over time to reflect the latest graph structure and interactions, facilitating the detection of new fraud patterns as they develop.

By leveraging dynamic graph modeling, GNNs can provide timely and accurate fraud detection, adapting to the evolving landscape of financial transactions.
\subsection{Integrating Multimodal Data}
Financial fraud detection often necessitates the analysis of multimodal data, which includes numerical transaction data, textual communication records, and categorical user attributes. The integration of such varied data types presents challenges, mainly due to their heterogeneity—numerical data, text, and categorical information each require distinct preprocessing techniques and feature extraction methods. However, this diversity also offers significant opportunities; by combining these different types of data, a more comprehensive and nuanced view of financial activities can be obtained, leading to more effective fraud detection strategies \cite{ngai2011application, lahat2015multimodal}.

\noindent\textbf{Multimodal Data Fusion.}
GNNs emerge as a powerful tool for fusing different types of data within a unified framework. GNNs can naturally accommodate multimodal data by representing them as nodes in a graph, where the edges capture the relationships between various data points~\cite{cheng2022financial}. For instance, a GNN can model transactions as nodes, with edges representing the flow of money between accounts, while incorporating textual data through node attributes or separate text-associated nodes linked to relevant transactions or accounts \cite{wu2020comprehensive}.

To effectively fuse multimodal data, a GNN can employ an approach such as:
\begin{equation}
\bm{z}_i^{(l)} = \bm{W}^{(l)} \cdot \text{AGGREGATE}\left(\{\bm{h}_j^{(l)} : j \in \mathcal{N}(i) \}\right)
\end{equation}
\begin{equation}
\bm{h}_i^{(l+1)} = \sigma\left( \bm{z}_i^{(l)} + \bm{B}^{(l)} \cdot \bm{x}_i \right)
\end{equation}
where $\bm{h}_i^{(l)}$ is the feature vector of node $i$ at layer $l$, $\bm{W}^{(l)}$ and $\bm{B}^{(l)}$ are learnable parameters, $\sigma$ is a non-linear activation function, $\text{AGGREGATE}$ is a function that combines features from node $i$'s neighbors, and $\bm{x}_i$ represents the multimodal data associated with node $i$, potentially combining numerical, textual, and categorical information into a unified representation. This modeling allows GNNs to leverage the rich, interconnected data landscape of financial transactions, providing a holistic view of financial activities. Through the integration of multimodal data, GNNs can uncover complex patterns of fraudulent behavior that might be overlooked when analyzing data sources in isolation. For more methods of integrating multimodal information, please refer to section 5.2.3.

\section{GNN Design for Financial Fraud Detection}
\label{sec:feat_eng}
In this section, we discuss the critical elements in designing GNNs tailored for effective financial fraud detection. Key considerations include the selection of appropriate GNN architectures, meticulous feature selection, and strategies for managing the challenge of imbalanced data to ensure robustness and effectiveness in detecting fraudulent activities.

\subsection{Architectural Choices for GNN}
Selecting an optimal GNN architecture for financial fraud detection involves understanding the unique characteristics of financial data. Graph Convolutional Networks excel in structured data environments, capturing local transaction patterns effectively. Graph Attention Networks leverage attention mechanisms to emphasize important transactions or patterns, beneficial in scenarios with significant edge attributes. Graph Transformer Networks adeptly handle time-series data, tracking financial activities over periods. Heterogeneous Graph Neural Networks are designed for networks with diverse node and edge types, enabling comprehensive integration of various financial behaviors.

\subsection{Feature Selection}
\subsubsection{Node Feature Engineering}
Node feature engineering in financial contexts is crucial. It involves a strategic analysis and extraction of node attributes, extending beyond basic transaction details like amount, frequency, and timing \cite{ikeda2021new}. Advanced methodologies such as clustering for geographic transaction analysis \cite{carpio2022towards}, entropy measures for transaction diversity \cite{isik2010entropy}, and time-series analysis for cyclic patterns \cite{devaki2014credit} significantly enrich the model's dataset, enhancing its predictive power against fraud.
\subsubsection{Edge Features Integration}
Edge features are crucial in GNNs for enhancing fraud detection accuracy by representing dynamic relationships between nodes. Typically, these features are integrated as propagation weights to tailor information aggregation and highlight significant relationships, as explored by Wang et al., demonstrating its effectiveness in emphasizing critical transactional connections indicative of fraudulent activity \cite{wang2022generalizing}. Additionally, enhancing the message-passing mechanism by incorporating edge features with node data enriches interactions between nodes and captures complex relational patterns essential for detecting fraud, as shown by Gong et al. \cite{gong2019exploiting}. Preprocessing edge features before feeding them into the GNN, as highlighted by Zheng et al., creates enriched node or edge attributes, enhancing the initial data representation and overall learning efficacy, crucial for uncovering rare fraudulent patterns that might otherwise be missed by more straightforward models \cite{zheng2024edge}. These techniques ensure that GNNs not only learn from but also emphasize the transactional connections most indicative of fraud, thereby enhancing the detection capabilities of financial fraud detection systems.

\subsubsection{Multimodal Data}
Integrating multimodal data such as text, images, and transaction logs is essential in enhancing GNNs for robust financial fraud detection. Multimodal data provides a richer representation of transactional contexts, enabling GNNs to detect complex fraudulent patterns effectively~\cite{cheng2022financial}.

\noindent\textbf{Self-Attention Weighted Fusion.} One effective technique is the use of self-attention weighted fusion, which dynamically adjusts the weights of different modal features to optimize data integration. Jia et al. developed a multi-modality self-attention aware deep network, originally designed for biomedical segmentation, demonstrating the effectiveness of this approach in handling complex, multimodal datasets \cite{jia2020multi}.

\noindent\textbf{Multimodal Embedding Fusion Layers.} Another approach is the use of multimodal embedding fusion layers that perform deep nonlinear transformations to merge data from various sources seamlessly. Kim and Chi's SAFFNet employs this technique for remote sensing scene classification, highlighting its potential to enhance feature integration across different modalities for improved classification accuracy \cite{kim2021saffnet}.

\noindent\textbf{Graph Convolution Interactions.} Integrating graph structures with multimodal data through graph convolution interactions enhances models' analytical capabilities. Zhang et al. employed this to predict urban dynamics, illustrating its effectiveness in spatiotemporal forecasting \cite{zhang2022multi}. Wei et al. developed MMGCN for personalized recommendations, showing its utility with diverse data types \cite{wei2019mmgcn}. Lian et al. demonstrated its application in completing graphs for conversation systems, ensuring comprehensive data utilization \cite{lian2023gcnet}.

These advancements highlight the importance of multimodal data integration in GNNs, significantly extending their applicability and effectiveness in financial fraud detection by providing a more comprehensive analysis of transactions across different modalities.

\subsection{Imbalanced Data}
The prevalence of imbalanced data in financial fraud detection, where legitimate transactions vastly outnumber fraud cases, presents a significant challenge. Several strategies have been developed to address this issue, enhancing the model's performance in detecting minority class instances.

\noindent\textbf{Graph Data Augmentation.} Techniques such as node sampling or edge manipulation can effectively rebalance the dataset. Zhao et al. introduced GraphSMOTE, which innovatively applies the SMOTE technique for imbalanced node classification on graphs \cite{zhao2021graphsmote}. Similarly, Perez-Ortiz et al. discussed graph-based over-sampling methods tailored for ordinal regression, which can be adapted for fraud detection \cite{perez2014graph}. Wu et al. developed a GNN-based fraud detector that specifically tackles the challenges posed by graph disassortativity and imbalance in data \cite{wu2024gnn}.

\noindent\textbf{Ensemble Methods in Graph Learning.} Aggregating predictions from various GNN models can improve accuracy on rare fraud cases. Pfeifer et al. explored the potential of federated ensemble learning with GNNs in the context of disease module discovery, which is analogous to fraud detection in its requirement for precise anomaly detection \cite{pfeifer2023ensemble}. Shi et al. proposed a boosting algorithm for GNNs that effectively addresses imbalanced node classification \cite{shi2021boosting}.

\noindent\textbf{Cost-sensitive GNN.} Incorporating cost-sensitive learning approaches in GNNs involves adjusting the loss function to emphasize correct predictions on the minority class. Hu et al. designed a cost-sensitive GNN specifically for mobile social network fraud detection, increasing penalties for misclassifying fraud instances \cite{hu2023cost}. Duan et al. and Ma et al. also developed dual cost-sensitive and attention mechanisms, respectively, for imbalanced node classification \cite{duan2022dual, ma2022attention}.

\noindent\textbf{Adaptive Resampling in GNN.} This strategy involves adjusting sampling rates based on current model performance to ensure balanced learning~\cite{li2024relation}. Zhang et al. detailed a hierarchical graph transformer with adaptive node sampling, optimizing sampling processes in real-time \cite{zhang2022hierarchical}. Hu et al. presented an adaptive two-layer light field compression scheme using a reconstruction method, which can be adapted for learning from imbalanced datasets \cite{hu2020adaptive}.

\noindent\textbf{Meta Paths in Heterogeneous Graph Networks.} Utilizing meta paths in heterogeneous graphs enhances the detection capabilities for nuanced fraudulent activities by capturing complex semantic information~\cite{li2024relation}. Liang et al. discussed the use of meta-path-based heterogeneous graph neural networks in academic networks, which can be adapted for fraud detection \cite{liang2022meta}. Wang et al. developed a heterogeneous graph attention network that leverages meta paths to focus on important relationships within the graph \cite{wang2019heterogeneous}. Meng et al. explored discovering meta-paths in large heterogeneous information networks, highlighting their importance in enhancing model learning capabilities \cite{meng2015discovering}.

These strategies collectively ensure that GNNs are optimized to handle the complexities of financial data, providing robust solutions against financial fraud.

\section{Application in Financial Fraud Detection}
GNNs have revolutionized the detection of fraudulent activities within diverse financial sectors. This section delves into the deployment of GNNs in key areas such as credit card fraud, online payment fraud, insurance fraud, and anti-money laundering efforts.

\subsection{Credit Card Fraud Detection}
Credit card fraud detection represents a significant challenge within the financial industry, focusing on identifying unauthorized transactions to prevent financial losses. The adoption of GNNs has introduced innovative methods to address this issue by leveraging relational data among transactions, cardholders, and merchants. By modeling these elements as nodes and relationships as edges within a graph, GNNs capture both explicit and subtle fraudulent patterns effectively.

Significant contributions include the attribute-driven graph representation by Xiang et al. \cite{xiang2023semi} within a semi-supervised learning framework, and the spatial-temporal attention mechanism by Cheng et al. \cite{cheng2020graph}, which significantly enhances fraud detection accuracy. Nguyen et al. \cite{dastidar2022importance} emphasize the predictive power of future transaction data, while Van Belle et al. \cite{van2022inductive} introduce an inductive learning approach that improves generalization to new instances of fraud.

\subsection{Online Payment Fraud Detection}
The detection of online payment fraud is critical for safeguarding digital financial transactions against various fraud types. GNNs, with their ability to model complex, dynamic relationships between transaction entities, have become a cornerstone in this domain. Notable developments include adaptive fraud detection frameworks in dynamic e-commerce environments by Zhang et al. \cite{zhang2022efraudcom}, and dual-level learning architectures that utilize transactional layers to enhance detection accuracy by Zheng et al. \cite{zheng2023midlg}. Hierarchical attention mechanisms, as explored by Hu et al. \cite{hu2019cash}, prioritize significant transactional features indicative of fraud. The application of heterogeneous graph models, such as those detailed by Liu et al. \cite{liu2018heterogeneous,liu2021intention}, effectively address multi-type relational data challenges, particularly in blockchain transactions.

\subsection{Insurance Fraud Detection}
In the insurance sector, GNNs analyze relational data among claims, providers, and patients to identify and predict fraudulent patterns. A notable application is Medicare fraud detection, where GNNs highlight inconsistencies in provider-patient interactions \cite{yoo2022medicare}. Techniques like risk diffusion in parallel GNN architectures address challenges associated with organized fraud groups, as noted by \cite{ma2023fighting}. Dynamic GNNs, which adapt to evolving schemes of collaborative fraud, are also crucial for continuous learning in fraud detection \cite{ren2023dynamic,zhang2022hierarchical}. These advancements demonstrate the effectiveness of GNNs in uncovering and mitigating insurance fraud.

\subsection{Anti-Money Laundering Detection}
Anti-money laundering is a critical challenge in the financial sector, aimed at identifying and curbing illicit financial activities to obscure the origins of criminal proceeds. Graph neural networks have introduced transformative methods by modeling accounts, transactions, and entities as nodes and their relationships as edges in a graph. This approach enables GNNs to detect complex laundering patterns effectively.

Significant contributions to this field include the application of self-supervised graph representation learning strategies, as explored by Cardoso et al. \cite{cardoso2022laundrograph}, which enhance the model's ability to detect unusual patterns autonomously. Additionally, Cheng et al. \cite{cheng2023anti} introduce group-aware deep graph learning approaches that improve the detection of coordinated laundering activities across multiple accounts. Furthermore, applications in cryptocurrency transactions have shown the potential of GNNs in financial forensics, with studies like those by Weber et al. \cite{weber2019anti} and Li et al. \cite{li2023graph} experimenting with GNNs to tackle money laundering in Bitcoin transactions.

Overall, GNNs offer a transformative approach to detecting and predicting fraudulent activities, outperforming traditional detection methods by leveraging the connectivity inherent in financial transaction data.

\section{Future Directions}
GNNs have shown remarkable potential for processing complex network data and enhancing fraud detection capabilities. As the field evolves, the following four research directions are pivotal for advancing the application of GNNs:

\noindent\textbf{Integration of Reinforcement Learning with GNNs.} As financial markets continue to evolve, there is a compelling need for models that can dynamically adjust and optimize in real-time. Reinforcement learning (RL) integrated with GNNs presents a promising approach to meet this demand. Initial studies, such as those by Munikoti et al. \cite{munikoti2023challenges} and Kong et al. \cite{kong2024mag}, have shown that RL can enhance the adaptability of GNNs, enabling them to effectively respond to changing transaction patterns and detect fraudulent activities more efficiently. Further investigations explore the diversity of approaches in enhancing GNNs with RL for fraud detection, demonstrating various strategies to optimize graph-based models for dynamic and accurate financial transaction analysis \cite{dou2020enhancing, chen2024scn_gnn, huang2022auc, li2022internet, jiang2021mafi}.
Future research should focus on developing models that further refine these adaptive capabilities, optimizing for real-time processing and decision-making across various financial scenarios. This could lead to more effective prediction and prevention of fraudulent activities by enabling continuous learning from transactional changes.

\noindent\textbf{Application of Adversarial Learning in GNNs.} The robustness of GNNs is critical in safeguarding against adversarial attacks, which simulate potential fraudulent activities. Adversarial learning enhances the accuracy and resilience of these models by training them to withstand various adversarial scenarios. Notable work in this area includes defending GNNs against adversarial attacks \cite{zhang2020gnnguard} and achieving certified robustness against structural perturbations \cite{wang2021certified}. Further studies such as those by Boyaci et al. \cite{boyaci2021graph} and Sun et al. \cite{sun2022adversarial} have extended this approach to different domains, including smart grids and general graph data systems. Future research should aim to develop more sophisticated adversarial training frameworks that expose GNNs to a broader array of attack scenarios, enhancing their efficacy and security in practical applications \cite{singh2021temporal,deng2022contrastive}.

\noindent\textbf{Pre-training in GNNs.} Inspired by the successes of large pre-trained models in natural language processing, similar strategies could be adopted for GNNs to advance graph data processing in finance. This approach has shown potential in uncovering deeper patterns and improving fraud detection accuracy. For instance, Hu et al. \cite{hu2019strategies} and Qiu et al. \cite{qiu2020gcc} have developed strategies and frameworks for pre-training GNNs on large and complex graph datasets. These techniques allow GNNs to learn useful representations before fine-tuning on specific tasks, enhancing performance across various domains. Further research should explore optimizing these models' training processes \cite{lin2020pagraph,yu2022graphfm,bai2021efficient} and enhancing feature extraction from large datasets to effectively combat increasingly sophisticated fraud schemes \cite{najafabadi2015deep,west2016intelligent}.

\noindent\textbf{Enhancing the Explainability of GNNs.}
The explainability of decision-making models is critical for gaining trust from regulators and users \cite{rao2021know,qin2022explainable,dai2021towards}. Future research should focus on developing methods or tools to elucidate GNNs' decision processes, aiming for greater transparency and understandability. Techniques could include advanced visualization methods to demonstrate how networks detect and react to fraudulent activities \cite{pandit2007netprobe,huang2009visualization}, or algorithms to clarify the significance of specific nodes and edges \cite{foster2010edge,meng2022novel}.

\noindent\noindent\textbf{Enhancing the Scalability of GNNs.}
Real-world financial networks, such as inter-bank transfer networks, often scale to millions of entities. This does not even account for the billions of historical unlabelled data points. Current single-machine single-GPU solutions can handle networks composed of up to ten million transactions~\cite{xiang2023semi}. Future research directions should consider leveraging multi-GPU setups to improve both training and inference efficiency. Additionally, utilizing disk space can increase the maximum size of transaction networks that can be processed. Furthermore, distributed computing frameworks should be considered to tackle billion scale fraud detection tasks using GNNs.

Exploring these directions will not only expand the application scope of GNNs in financial fraud detection but also provide a robust theoretical and practical foundation for the ongoing development of these technologies.
\section{Conclusion}
This study evaluates the efficacy of GNNs in the detection of financial fraud by addressing a set of comprehensive research questions that encompass the methodologies, roles, design, deployment, and challenges associated with GNNs within the financial fraud detection field. The introduction of a unified framework helps to systematize the diverse approaches observed in existing literature, offering a clearer understanding of how GNNs can be effectively applied in complex financial contexts. Our findings contribute to the body of knowledge by detailing the adaptive capabilities of GNNs to the dynamic nature of financial networks and emphasizing their potential to enhance fraud detection mechanisms. Moreover, this work identifies existing gaps and outlines prospective research directions that could further solidify the role of GNNs in improving fraud prevention systems. The insights provided here aim to guide future studies toward more nuanced applications and integration strategies for GNNs in real-world financial systems.

\begin{acknowledgement}
This work was supported by the National Key R\&D Program of China (Grant no. 2022YFB4501704), the National Natural Science Foundation of China (Grant no. 62102287), and the Shanghai Science and Technology Innovation Action Plan Project (Grant no. 22YS1400600 and 22511100700).
\end{acknowledgement}
\bibliographystyle{fcs}
\bibliography{ref}

\end{document}